\documentclass{article}
\usepackage{spconf,amsmath,graphicx}

\usepackage{caption} 
\usepackage{tabularx}
\usepackage{float}
\usepackage{array}  
\usepackage{booktabs} 
\usepackage{multirow}
\usepackage{makecell}

\usepackage{color}

\usepackage{bbding}

\usepackage{algorithm}
\usepackage{algorithmic} 

\usepackage{setspace}

\usepackage{arydshln} 

\definecolor{green(ryb)}{rgb}{0.4, 0.69, 0.2}
\definecolor{ao(english)}{rgb}{0.0, 0.5, 0.0}

\usepackage{amsfonts,amssymb} 
\usepackage{bm} 

\usepackage{subfig} 

\usepackage{xspace}
\newcommand{\event}[1]{\textit{\footnotesize #1}\xspace}
\renewcommand{\S}{\event{\textless S\textgreater}}
\newcommand{\Sprime}{\event{\textless S$'$\textgreater}}
\newcommand{\E}{\event{\textless E\textgreater}}
\newcommand{\esub}[1]{\textit{\footnotesize event}\textsubscript{#1}\xspace}

\title{GCT: Gated Contextual Transformer for Sequential Audio Tagging}

\name{Yuanbo Hou$^{1}$, Yun Wang$^2$, Wenwu Wang$^3$, Dick Botteldooren$^{1}$}
 \address{
 $^1$WAVES, Ghent University, Belgium  \quad $^2$Meta AI, USA  \quad $^3$CVSSP, University of Surrey, UK }

\begin{document}


\maketitle
\begin{abstract}
\vspace{-0.05cm}
Audio tagging aims to assign predefined tags to audio clips to indicate the class information of audio events. 
Sequential audio tagging (SAT) means detecting both the class information of audio events, and the order in which they occur within the audio clip. 
Most existing methods for SAT are based on connectionist temporal classification (CTC). However, CTC cannot effectively capture connections between events due to the conditional independence assumption between outputs at different times.
The contextual Transformer (cTransformer) addresses this issue by exploiting contextual information in SAT.  
Nevertheless, cTransformer is also limited in exploiting contextual information as it only uses forward information in inference. 
This paper proposes a gated contextual Transformer (GCT) with forward-backward inference (FBI). 
In addition, a gated contextual multi-layer perceptron (GCMLP) block is proposed in GCT to improve the performance of cTransformer structurally.
Experiments on two real-life audio datasets show that the proposed GCT with GCMLP and FBI performs better than the CTC-based methods and cTransformer.
To promote research on SAT, the manually annotated sequential labels for the two datasets are released.

\end{abstract}
\vspace{-0.05cm}

\vspace{-0.25cm}
\section{Introduction}\label{sec:intro}
\vspace{-0.2cm}
Audio tagging (AT)~\cite{audio_cl} is a fundamental task in audio classification, where events in audio clips are identified via multi-label classification. 
Sequential audio tagging (SAT)~\cite{ctsat} aims to mine both the type information of events and the order information between events.
With SAT, the number of events can be estimated as a byproduct.
Sequence-level AT offers more information about the audio clip than temporal agnostic event-level AT. 
SAT can not only provide information that is useful for AT tasks \cite{audio_cl}, but also support tasks like audio captioning \cite{audio_caption}, event anticipation \cite{event_pred} and scene perception \cite{acoustic_scene}.

\begin{figure*}[t]
\centering
\captionsetup[subfigure]{labelformat=empty} 
\subfloat[]{\includegraphics[width=5.3in]{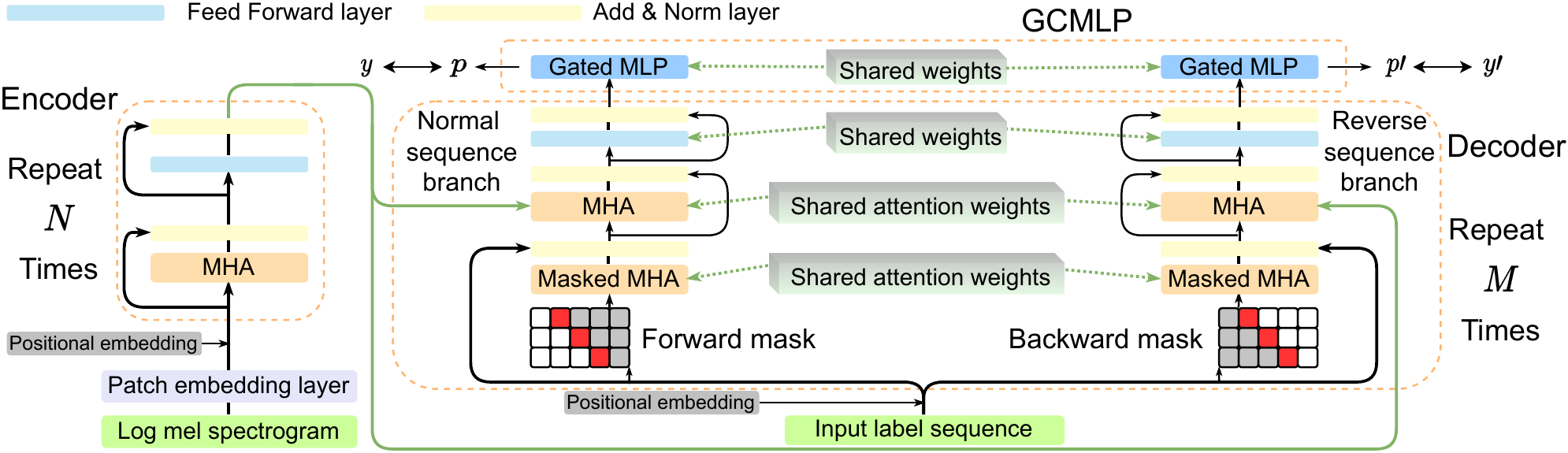}}
\hspace{4mm} 
\subfloat[]{\includegraphics[width=1in]{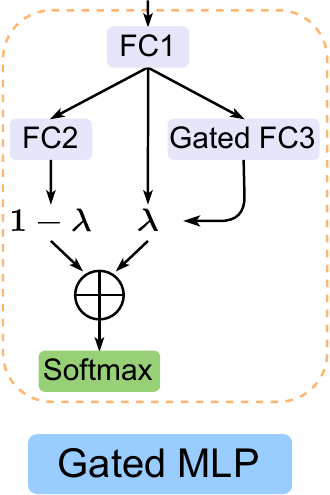}}
\setlength{\abovecaptionskip}{-0.4cm}  
\setlength{\belowcaptionskip}{-0.6cm}  
\caption{\small{The proposed gated contextual Transformer and gated MLP block. 
In the mask matrices, the red, gray, and white blocks present the positions corresponding to the target to be predicted, the positions of masked data, and the positions of available data.}}
\label{model}
\end{figure*}

Prior works on SAT mostly use connectionist temporal classification (CTC) \cite{ctc}
as its core.
To perform SAT on polyphonic audio clips, sequential labels are introduced in \cite{dcase2018_ctc} to train CTC-based convolutional recurrent neural networks to tag diverse event sequences.
The polyphony of audio makes it hard to define the order of events, therefore, the order of the beginning and end boundaries of events are used as sequential labels in \cite{dcase2018_ctc}. 
The double-boundary sequential labels are also used to train a CTC-based recurrent neural network equipped with the long short-term memory (BLSTM-CTC) \cite{wangyun} to perform sound event detection (SED), which detects the type and temporal position of events in audio clips. 
Apart from these methods using double-boundary labels, single-boundary sequential labels (sequences of the start boundary of events) are exploited in a 2-stage CTC-based method \cite{hou2019sound} for SED.
However, CTC-based methods have difficulty in modeling the contextual information in event sequences because CTC implicitly assumes that the network outputs are conditionally independent at each time step \cite{ctc}.
To take advantage of context, a contextual Transformer (cTransformer) \cite{ctsat} is proposed to explore bidirectional information in event sequences to make more effective use of the contextual information in SAT.

\vspace{-0.1cm}
To exploit both the forward and backward information of events, cTransformer uses a bidirectional decoder to model the correlations between preceding and following events in both directions in training, while only the forward direction of the decoder is used for inferring \cite{ctsat}. In inference, cTransformer does not utilize the event sequences information contained in the reverse sequence branch. To address this limitation, we propose a gated contextual Transformer (GCT) with a forward-backward inference (FBI) algorithm to infer the target event from both directions, where the context of the event is incorporated during inference. In addition, to enhance the decoder's power to capture the context implicit in event sequences, a gated contextual multi-layer perceptron (GCMLP) block is proposed to adapt the contextual information for the estimation of final predictions.

\vspace{-0.1cm}
The contributions of this work are: 
1) We propose GCT equipped with GCMLP and FBI to improve cTransformer's ability to capture contextual information of event sequences;
2) We explore the effect of pretrained weights on modules of GCT under two transfer learning modes to gain insight into the role of GCT modules;
3) We visualize the attention distribution in hidden layers of GCT to investigate how GCT connects acoustic representations with clip-level event tokens and bidirectionally infers the event sequences;
4) To evaluate the performance of GCT, we sequentially annotate a polyphonic audio dataset and release it. We compare the performance of GCT, cTransformer, and CTC-based methods on two real-life datasets. 
This paper is organized as follows, Section 2 introduces the GCT. Section 3 describes the dataset, experimental setup, and analyzes the results. Section 4 gives conclusions.

\section{Gated contextual Transformer (GCT)}\label{sec:gct}

\vspace{-0.325cm} 
Following the approach in cTransformer \cite{ctsat}, 
the sequence of event start boundaries is used as the sequential label.
For example,  given the normal sequential label $l$ is ``\S, \esub{1}, \esub{2}, ..., \esub{$k$}, \E'', where $k$ is the number of events, \S and \E are the tokens indicating the start and end of the sequence, respectively.
The reverse sequential label $l^\prime$ is ``\Sprime, \esub{$k$}, \esub{$k$-1}, ..., \esub{1}, \E'', where \Sprime is the token indicating the start of the reverse sequence.

\vspace{-0.35cm}
\subsection{Encoder and Decoder of GCT}

\vspace{-0.2cm}
\textbf{Encoder.}
There are two ways for the input to the encoder: 1) the entire spectrogram of the audio clip, as in cTransformer \cite{ctsat}; 2) the patch sequence by dividing the spectrogram clip into patches, as in AST \cite{ast}. 
Inputting the entire clip enables the encoder to directly utilize the global audio information of events. However, the patch sequence may help the model to align acoustic patch sequences with the corresponding event label sequences.   
Fig.~\ref{model} shows the structure of GCT with input patches.
Referring to AST, GCT uses a patch embedding layer to map the patches containing basic acoustic features into high-level representations, and uses updatable positional embedding (Pos\_emb) to capture the spatial information of each patch.
When inputting clips, the Pos\_emb before the encoder is removed, and the patch embedding layer is replaced with a linear layer to keep the encoder input dimension consistent with input patches.
The encoder consists of $N$ identical blocks with a multi-head attention layer (MHA) and a feed forward layer (FF) with layer normalization, which are analogous to the encoder in Transformer \cite{Transformer}, so all the parameters in MHA and FF follow Transformer's default setting.

\textbf{Decoder.}
The bidirectional decoder in GCT consists of a normal sequence branch and a reverse sequence branch.  
To facilitate information exchange between branches, in Fig.~\ref{model}, each branch consists of $M$ identical blocks serially, and each block consists of similar structures containing masked MHA, MHA, and FF. 
To preserve the autoregressive property \cite{Transformer} of Transformer-based models, forward and backward masks are used to block future and past information about the target in the normal and reverse branches, respectively, to maintain the sequential information between events in a sequence.
With the combined effect of the forward and backward mask matrices, the normal and reverse sequence branches will infer the same target at each time step. Thus, the decoder can extract forward and backward information about the target.
Furthermore, the weights shared between branches will facilitate the model to capture contextual information about the target.

\vspace{-0.35cm}
\subsection{Gated contextual multi-layer perceptron (GCMLP)}

\vspace{-0.2cm}
GCMLP aims to perform the final conditioning of the decoder output based on the gated MLP (gMLP) block and shared weights while considering the contextual information about the target to achieve more accurate predictions.
In Fig.~\ref{model}, gMLP consists of 3 fully-connected (FC) layers with the same size. 
Denote the input as $X$, the weight of FC1, FC2 and FC3 as $W_1$, $W_2$ and $W_3$, then the corresponding output is $F_1 = W_1 X$, $F_2 = ReLU(W_2 {F}_{1})$, and $\lambda= \sigma(W_3 {F}_{1})$, where  $\sigma$ is logistic sigmoid: $\sigma(x) = 1/({1+e^{-x}})$, and $ReLU$ is activation function \cite{ReLU}. Then, the output of gMLP is  
\begin{equation}
\setlength{\abovedisplayskip}{1pt}
\setlength{\belowdisplayskip}{1pt} 
gMLP = Softmax((1-\lambda)\odot F_2 + \lambda\odot F_1)
\end{equation} 
Where $\odot$ is the element-wise product, 
$F_2$ is a higher-level representation of the target based on $F_1$, and can be viewed as the target's embedding from another perspective. 
FC3 evaluates the relative importance of each element in $F_1$, and then combines it with $F_2$ according to the estimated importance of each element.
That is, gMLP is used to generate multi-view results and fuse them relying on the learnable gate unit.

At each time step, denote the output of the normal and reverse branches through GCMLP as $p$ and $p\prime$, the corresponding ground-truth labels are $y$ and $y\prime$, respectively. 
Cross entropy (CE) \cite{Transformer} loss is used in GCT: $\mathcal{L}_\text{normal}=CE(p, y)$, $\mathcal{L}_\text{reverse}=CE({p\prime}, {y\prime})$. 
To further align the classification space of the two branches to allow the model to focus on the contextual information of the same target, the mean squared error (MSE) \cite{mse} is used as a context loss to measure the distance between $p$ and $p\prime$ in the latent space: $\mathcal{L}_\text{context} = MSE({p\prime}, p)$.
Hence, the total loss of GCT is $\mathcal{L} = \mathcal{L}_\text{normal}+ \mathcal{L}_\text{reverse}+\mathcal{L}_\text{context}$.

\vspace{-0.375cm}
\subsection{Forward-backward inference (FBI)} 

\vspace{-0.2cm}
To utilize both the normal and reverse branches in inference, we propose FBI to make the two branches infer the same goal at each step, and fuse the prediction to form the final output. 
While preserving the autoregressive property \cite{Transformer}, FBI integrates the forward and the backward sequence information implied in the normal and reverse branches during inference, which benefits the model to make fuller use of the contextual information of events in inference.
Details of FBI are below.

%
\begin{algorithm}[H]
\setlength{\parskip}{0pt}
\footnotesize
	\caption{\small{PyTorch pseudo code for the proposed FBI}} \label{fbi}  
 {\color{ao(english)}{\# $\boldsymbol{X}$: input log mel spectrogram; $\boldsymbol{X'}$: $\boldsymbol{X}$ reversed along the time axis}}  
 
 $\boldsymbol{E}$, $\boldsymbol{E'}$ = Encoder($\boldsymbol{X}$), Encoder($\boldsymbol{X'}$)  \qquad\qquad\qquad\qquad{\color{ao(english)}{\# output of encoder}}

 $\boldsymbol{I}$, $\boldsymbol{I'}$ = \S, \Sprime \quad\quad\quad{\color{ao(english)}{\# start token of the normal and reverse sequence}}

 for $k$ in range($L - 1$): \quad{\color{ao(english)}{\# $L$: max length of event sequences; $B$: batch size}}
 
 \hphantom{1in}{$\boldsymbol{D}$ = Decoder\_normal\_branch(\boldsymbol{$E$}, $\boldsymbol{I}$)} \quad\quad{\color{ao(english)}{\# $\boldsymbol{D}$: ($B$, $L$, number of tokens)}} 

 \hphantom{1ex}{$\boldsymbol{p}$ = GCMLP($\boldsymbol{D}$[:, -1, :])} \quad\quad\quad{\color{ao(english)}{\# pick the latest target probability vector}}

\vspace{0.5em}

 \hphantom{1in}{\color{ao(english)}{\# \textbf{Key step}: the reverse sequence branch predicting reversed input is}}
 \hphantom{1in}{\color{ao(english)}{equivalent to the normal sequence branch predicting normal input.}}
 
 \hphantom{1in}{$\boldsymbol{D'}$ = Decoder\_reverse\_branch($\boldsymbol{E'}$, $\boldsymbol{I'}$)}

 \hphantom{1in}{$\boldsymbol{p\prime}$ = GCMLP($\boldsymbol{D'}$[:, -1, :])} \quad{\color{ao(english)}{\# $p\prime$ and $p$ are the same target's predictions}}

 \vspace{0.5em}

 \hphantom{1in}{$\boldsymbol{p_{ci}}$ = $\alpha \boldsymbol{p}$ + $(1-\alpha) \boldsymbol{p\prime}$
 \ {\color{ao(english)}{\# $\boldsymbol{p_{ci}}$: final prediction with contextual informa-}}
 \hphantom{1in}{\color{ao(english)}{mation; $\alpha$: importance factor of the forward information, default to 0.5.}}} 

 \hphantom{1in}{\_, $\boldsymbol{p_{et}}$ = torch.max($\boldsymbol{p_{ci}}$, dim=1).item()} \qquad{\color{ao(english)}{\# $\boldsymbol{p_{et}}$: predicted event token}} 

 \hphantom{1in}{if  $\boldsymbol{p_{et}}$ == \E: \textbf{break}} \  \  \ \  \qquad\qquad{\color{ao(english)}{\# \E: end token of event sequences}}

\vspace{0.5em}

 \hphantom{1in}{$\boldsymbol{I}$ = torch.cat([$\boldsymbol{I}$, torch.ones(1, 1).fill\_($\boldsymbol{p_{et}}$)], dim=1)}

 \hphantom{1in}{$\boldsymbol{I'}$ = torch.cat([$\boldsymbol{I'}$, torch.ones(1, 1).fill\_($\boldsymbol{p_{et}}$)], dim=1)}

\end{algorithm}

\vspace{-0.6cm}
\section{Experiments and results}

\begin{table}[b]\footnotesize 
	\setlength{\abovecaptionskip}{0cm}   
	\setlength{\belowcaptionskip}{-0.4cm}   
	\renewcommand\tabcolsep{1.1pt} 
	\centering 
\caption{\small{AUC of different input modes of GCT with different numbers of encoder and decoder blocks on the \textit{Noiseme} dataset.}} 
	\begin{tabular}
	{p{0.2cm}<{\centering}|
    p{0.28cm}<{\centering}
	p{0.28cm}<{\centering}|
	p{1.55cm}<{\raggedright}  
	p{1.55cm}<{\raggedleft}|
	p{0.25cm}<{\centering}|
	p{0.28cm}<{\centering}
	p{0.28cm}<{\centering}|
	p{1.55cm}<{\raggedright}
	p{1.5cm}<{\raggedleft}} 

\hline
		{\#} & $N$ & $M$ & \makecell[c]{\textsl{Patches}} & \makecell[c]{\textsl{Clip}} & {\#} & $N$ & $M$  & \makecell[c]{\textsl{Patches}} & \makecell[c]{\textsl{Clip}} \\ 
		
		\specialrule{0em}{0.1pt}{0.1pt}
		
        \hline
        \specialrule{0em}{0.1pt}{0.1pt}
  
		1 & 1 & 2 &  0.575$\pm$0.010 & 0.647$\pm$0.012
  & 7 & 8 & 6 &  0.534$\pm$0.033  &  0.557$\pm$0.058    \\ 
  
		2 & 2 & 4  & 0.584$\pm$0.009 & 0.661$\pm$0.016
  & 8 & 8 & 8 &  0.614$\pm$0.020  &  0.518$\pm$0.063   \\
		
		3 & 4 & 4 &  0.600$\pm$0.018   &  \textbf{0.662$\pm$0.013}  
  & 9 & 9 & 5 &  0.609$\pm$0.026  &  0.512$\pm$0.017  \\

  4 & 5 & 5 &  0.599$\pm$0.046 &   0.660$\pm$0.071 
  & 10 & 9 & 7 &  0.604$\pm$0.066  &  0.511$\pm$0.013  \\

  5 & 6 & 6 &  0.609$\pm$0.017  &  0.596$\pm$0.075 
  & 11 & 9 & 9 &  0.608$\pm$0.027  & 0.511$\pm$0.007   \\

  6 & 7 & 7 & \textbf{0.627$\pm$0.019}  &   0.543$\pm$0.024 	
  & 12 & 10 & 10 &  0.606$\pm$0.052  &  0.508$\pm$0.032  \\
		 
		\hline
	\end{tabular}
	\label{tab:input_blocks}
\end{table}

\vspace{-0.3cm}
\subsection{Dataset, Baseline, Experiments Setup, and Metrics}
\vspace{-0.1cm} 
The \textit{DCASE}~\cite{dcase2018} and \textit{Noiseme}~\cite{noiseme} real-life audio datasets are used.
\textit{DCASE} contains 10 classes of audio events, where the training and test sets consist of 1578 and 288 10-second (10s) audio clips, respectively.
\textit{Noiseme} contains 33 classes of audio events, where the training and test sets consist of 2312 and 505 10s audio clips, respectively. 
In training, 20\% of training samples are randomly selected to form the validation set.
The sequential labels of \textit{DCASE} are here~\cite{ctsat}. 
We manually annotated and released sequential labels of \textit{Noiseme}. 
\textit{Noiseme} offers advantages over \textit{DCASE} in overall duration and number of event classes, so \textit{Noiseme} is mainly used in experiments.

This paper uses BLSTM-CTC \cite{wangyun}, convolutional bidirectional gated recurrent units with CTC (CBGRU-CTC) \cite{csps_ctc}, 
CBGRU-CTC with GLU in convolutional layers (CGLU-BGRU-CTC) \cite{dcase2018_ctc} and in convolutional-recurrent layers (CBGRU-GLU-CTC) \cite{hou2019sound} as baselines, and also presents the performance of cTransformer~\cite{ctsat}, which is the first Transformer-based model to use sequential labels for SAT.

For acoustic features, log mel-bank energy with 128 banks \cite{mel} is used with a Hamming window length of 25\textit{ms} and a hop size of 10\textit{ms} between windows. 
A batch size of 64 and stochastic gradient descent with a momentum 0.99 \cite{sgdm} with an initial learning rate of 1e-3 are used to minimize the loss.
Layer normalization \cite{layernorm} and dropout \cite{dropout} are used to prevent overfitting. 
Systems are trained on a card Tesla V100 GPU for 500 epochs.
Accuracy (\textsl{Acc}), \textsl{F-score} \cite{metrics}, and area under curve (\textsl{AUC}) \cite{AUC} are used to quantify the classification ability of models on audio events.
The BLEU score ~\cite{bleu} is used to evaluate the accuracy of predicted audio event sequences.
For more details and the labeled dataset, please visit the homepage
{\footnotesize{(\textcolor{blue}{\underline{https://github.com/Yuanbo2020/GCT}})}}.

\vspace{-0.3cm}
\subsection{Results and Analysis}

\vspace{-0.2cm}
\textbf{Model structure.} 
The optimal configuration of GCT is explored in Table \ref{tab:input_blocks}. 
Changes in the number of blocks do not significantly affect the performance of GCT when the input is patches. 
In addition, shallow GCT ($N\leq 6$, $M\leq 6$)  outperforms deep GCT when the input is clips, probably because the manually labeled dataset is not large enough to take advantage of large and deep models.
Deep neural networks are hard to train \cite{model_training}. 
System \#12 in Table~\ref{tab:input_blocks} has about 170 layers, and the vanishing gradient \cite{vanishing} we observed during training implies that such a deep model is challenging to optimize effectively, which may also be a reason for the poor performance of deep GCT.
We will use \{7, 7\} and \{4, 4\} as default structures for GCT when the input is patches and clips, respectively.

\textbf{Ablation study.}
Prior work \cite{ctsat} shows the benefits of context for SAT, so this paper focuses on the role of GCT components.
GCT with patches has a more complex structure than GCT with spectrogram clips. 
With patches as input, GCT uses Pos\_emb to capture the position information of patches. 
In Table~\ref{tab:input_blocks}, GCT does not degrade much as the number of $N$ and $M$ are increased when patches equipped with Pos\_emb are input. Does this mean that Pos\_emb is a crucial component of GCT? Table~\ref{tab:ablation_component} conducts ablation experiments for this.

\begin{table}[H]\footnotesize   
	\setlength{\abovecaptionskip}{0cm}   
	\setlength{\belowcaptionskip}{-0.2cm}   
	\renewcommand\tabcolsep{0.7pt} 
	\centering
	\caption{\small{Ablation study of GCT \{7, 7\} component on \textit{Noiseme}.}}
	\begin{tabular}
	{p{0.4cm}<{\centering}|
	p{1.4cm}<{\centering}
	p{1.2cm}<{\centering}|
	p{1.8cm}<{\centering}
 p{1.5cm}<{\centering}
	p{1.8cm}<{\centering}  }
 \hline 
	\#  
 & \textsl{Pos\_emb}
 & \textsl{GCMLP}
   & \textsl{\textbf{AT:} Acc (\%)} 
    & \textsl{\textbf{AT:} AUC}
   & \textsl{\textbf{SAT:} BLEU} \\
		\hline  
		\specialrule{0em}{0.05em}{0.pt}

  1 & \XSolidBrush & \XSolidBrush  & 92.23$\pm$0.61 & 0.600$\pm$0.014 & 0.297$\pm$0.045  \\

		2 & \CheckmarkBold & \XSolidBrush   & 93.00$\pm$0.54 & 0.616$\pm$0.012 & 0.312$\pm$0.019   \\ 
  
		3 & \XSolidBrush & \CheckmarkBold   &  92.55$\pm$0.62 & 0.610$\pm$0.009 & 0.309$\pm$0.023  \\ 
  
  4 & \CheckmarkBold & \CheckmarkBold  & \textbf{93.21$\pm$0.27}  & \textbf{0.627$\pm$0.019} &  \textbf{0.338$\pm$0.012}  \\ 
 
	\hline
	\end{tabular}
	\label{tab:ablation_component}
\end{table}
\vspace{-0.2cm}
In Table~\ref{tab:ablation_component}, Pos\_emb (shown as \#2) slightly outperforms GCMLP (i.e. \#3). This reveals that when the input is small patches, the position information is valuable for the model to effectively capture the local information of events. 
Combining Pos\_emb and GCMLP offers better results, shown as \#4 in Table~\ref{tab:ablation_component}, due to the use of local context information of events. 

Table~\ref{tab:fbi_method} illustrates the role of the proposed FBI. 
Overall, FBI plays a more powerful role when coarse-grained clips are input. 
The reason may be that after the spectrogram is split into patches, the time interval between forward and reverse information is shortened in each fine-grained patch, equivalent to reducing the range of context that FBI can capture. Hence, FBI is more valuable for clips as input than patches.

\begin{table}[b]\footnotesize 
	\setlength{\abovecaptionskip}{0cm}   
	\setlength{\belowcaptionskip}{-0.4cm}   
	\renewcommand\tabcolsep{0.9pt} 
	\centering 
\caption{\small{Ablation study of the inference method on \textit{Noiseme}.}} 
	\begin{tabular}
	{p{0.3cm}<{\centering}|
    p{0.58cm}<{\centering} |
	p{1.55cm}<{\centering}  
	p{1.55cm}<{\centering}|
	p{0.3cm}<{\centering}|
	p{0.58cm}<{\centering}|
	p{1.55cm}<{\centering}
	p{1.5cm}<{\centering}}  

\hline 
		\multirow{3}{*}{\makecell[c]{\rotatebox{90}{Acc (\%)}}} 
  & \textsl{FBI} & \makecell[c]{\textsl{Patches}} & \makecell[c]{\textsl{Clip}} 
   & \multirow{3}{*}{\makecell[c]{\rotatebox{90}{AUC}}} & \textsl{FBI}  & \makecell[c]{\textsl{Patches}} & \makecell[c]{\textsl{Clip}} \\ 
		
		\specialrule{0em}{0.1pt}{0.1pt}
        \cline{2-4} 
        \cline{6-8}
        \specialrule{0em}{0.1pt}{0.1pt}

		  &  \XSolidBrush  &   93.21$\pm$0.27 & 93.49$\pm$0.39
  &   &  \XSolidBrush &  0.627$\pm$0.019  &  0.662$\pm$0.013    \\

&  \CheckmarkBold  & {93.57$\pm$0.46} &  \textbf{94.01$\pm$0.31}    
  &   &  \CheckmarkBold &  {0.635$\pm$0.014}  &  \textbf{0.685$\pm$0.022}  \\		 
		\hline
	\end{tabular}
	\label{tab:fbi_method}
\end{table} 

\textbf{Pretrained weight.}
Table~\ref{tab:ablation_component} indicates the role of Pos\_emb for GCT with patches as input. 
In GCT, both Pos\_emb and encoder designs are inspired by AST~\cite{ast}.
Compared with datasets used in this paper, the 5800-hour Audioset \cite{audioset} used to train AST is super large-scale. 
If the knowledge learned from Audioset by AST is transferred to GCT, what impact will it have on GCT?
Table~\ref{tab:pretraining} shows the results of transferring parameters of AST to the corresponding parts of GCT in the fixed and fine-tuning modes \cite{hou2020transfer}. 
In training, the rest of GCT is randomly initialized, except for the part that uses transferred parameters.
To make the result of GCT a benchmark for Transformer-based models on \textit{DCASE} dataset, in Table~\ref{tab:pretraining}, $\{N, M\}$ = $\{6, 6\}$ of Transformer \cite{Transformer} is adopted in GCT.

\begin{table}[b]\footnotesize   
	\setlength{\abovecaptionskip}{0cm}   
	\setlength{\belowcaptionskip}{-0.55cm}  
	\centering
	\caption{\small{Effect of transfer learning on GCT on \textit{DCASE}.}}
 \begin{tabular}
	{p{0.3cm}<{\centering}|
	p{1.3cm}<{\centering}
	p{1.3cm}<{\centering}|
	p{1.6cm}<{\centering}
	p{1.6cm}<{\centering}}

 \hline 

 	\#
 &  \textsl{\textbf{Pos\_emb}} & \textsl{\textbf{Encoder}}  & \textsl{\textbf{AT:} Acc (\%)} 
 & \textsl{\textbf{SAT:} BLEU} 
 \\
 \hline
   
  1 &  \multicolumn{2}{c|}{No Transfer}  & 89.13$\pm$0.58 &  0.435$\pm$0.037       \\

		2 &  Fixed & Fixed  & \textbf{97.68$\pm$0.18} &  \textbf{0.677$\pm$0.014}    \\
  3 &  Fine-tuned & Fine-tuned  & 96.27$\pm$0.36 &  0.645$\pm$0.019      \\
  4 &  Fixed & Fine-tuned  & 93.84$\pm$0.85 &  0.639$\pm$0.016       \\
  5 &  Fine-tuned & Fixed  & 96.45$\pm$0.47 &  0.662$\pm$0.015       \\
	\hline
	\end{tabular}
	\label{tab:pretraining}
\end{table}

Compared with \#1 in Table~\ref{tab:pretraining}, which is randomly initialized and trained, the remaining models using pretrained weights achieve larger improvements, which proves the benefit of knowledge from large datasets in improving models' performance.
Specifically, \#5 outperforms \#4, indicating that the encoder with the ability in acoustic feature extraction is more important than Pos\_emb in providing the position information of patches.  
The fixed mode (i.e. \#2) is better than fine-tuning the transferred parameters (i.e. \#3).
The reason may be that the part (Pos\_emb and encoder) containing pretrained weights and the remaining randomly initialized part (decoder and GCMLP) differ greatly in the latent space, 
fine-tuning these two disparate parts using the same learning rate will inevitably affect the performance of (Pos\_emb and encoder) with audio events expertise. 
Freezing the parameters containing audio knowledge in \#2 will reduce the learning burden of GCT, which helps GCT to focus more on optimizing the remaining parts, thus achieving better results.

\textbf{Comparison with prior methods.}
In Table~\ref{tab:other_models}, cTransformer and GCT are multi-loss models, and tuning the ratio among losses can lead to better scores. 
In this paper, the weights of the sub-losses of GCT default to 1. 
To compare fairly, the weights of cTransformer losses are also set to 1. 
For single event recognition (i.e. AT),  CTC-based models perform similarly to Transformer-based models. 
But for modeling of event sequences (i.e. SAT), Transformer-based models work better.  
That means CTC, which assumes that networks' outputs at different times are conditionally independent, is inferior to Transformer in its ability to model long-term dependences on sequences.
Compared to Transformer, cTransformer performs better because it captures contextual information of events in sequences. 
The proposed GCT, which improves cTransformer in structure and inference, achieves the best results on AT and SAT tasks.

\begin{table}[b]\footnotesize
	\setlength{\abovecaptionskip}{0cm}   
	\setlength{\belowcaptionskip}{-0.5cm}   
	\renewcommand\tabcolsep{0.8pt} 
	\centering 
	\caption{\small{Comparison of other methods on \textit{Noiseme}.}}
	\begin{tabular}{
	p{2.8cm}<{\centering}|
	p{1.9cm}<{\centering}
	p{1.7cm}<{\centering}
	p{1.8cm}<{\centering}} 
		\hline 
	{Method} & \textsl{\textbf{AT:} F-score (\%)} & \textsl{\textbf{AT:} AUC} & \textsl{\textbf{SAT:} BLEU}\\
		\hline 
		\specialrule{0em}{0.07em}{0.pt}
		BLSTM-CTC \cite{wangyun} & 44.06$\pm$1.92  & 0.549$\pm$0.021 &  0.252$\pm$0.047 \\

		CBGRU-CTC \cite{csps_ctc} & 48.55$\pm$1.74  & 0.583$\pm$0.009 &  0.290$\pm$0.043 \\

		CGLU-BGRU-CTC \cite{dcase2018_ctc} & 50.03$\pm$1.66  & 0.579$\pm$0.011 &  0.297$\pm$0.015 \\
		
		CBGRU-GLU-CTC \cite{hou2019sound} & 48.79$\pm$1.87  & 0.572$\pm$0.007 &  0.275$\pm$0.011 \\

		Transformer \cite{Transformer} & 45.98$\pm$0.75  & 0.563$\pm$0.027 &  0.310$\pm$0.019 \\

		cTransformer \cite{ctsat} &  46.44$\pm$1.10  & 0.577$\pm$0.032 &  0.323$\pm$0.009 \\

    Proposed GCT (Clip) & \textbf{56.03$\pm$1.53}  & \textbf{0.685$\pm$0.022} & \textbf{0.403$\pm$0.068} \\

		\hline 
	\end{tabular}
	\label{tab:other_models}
\end{table}

\begin{figure}[t] 
	\setlength{\abovecaptionskip}{0cm}  
	\setlength{\belowcaptionskip}{-0.5cm}   
	\centerline{\includegraphics[width = 0.5 \textwidth]{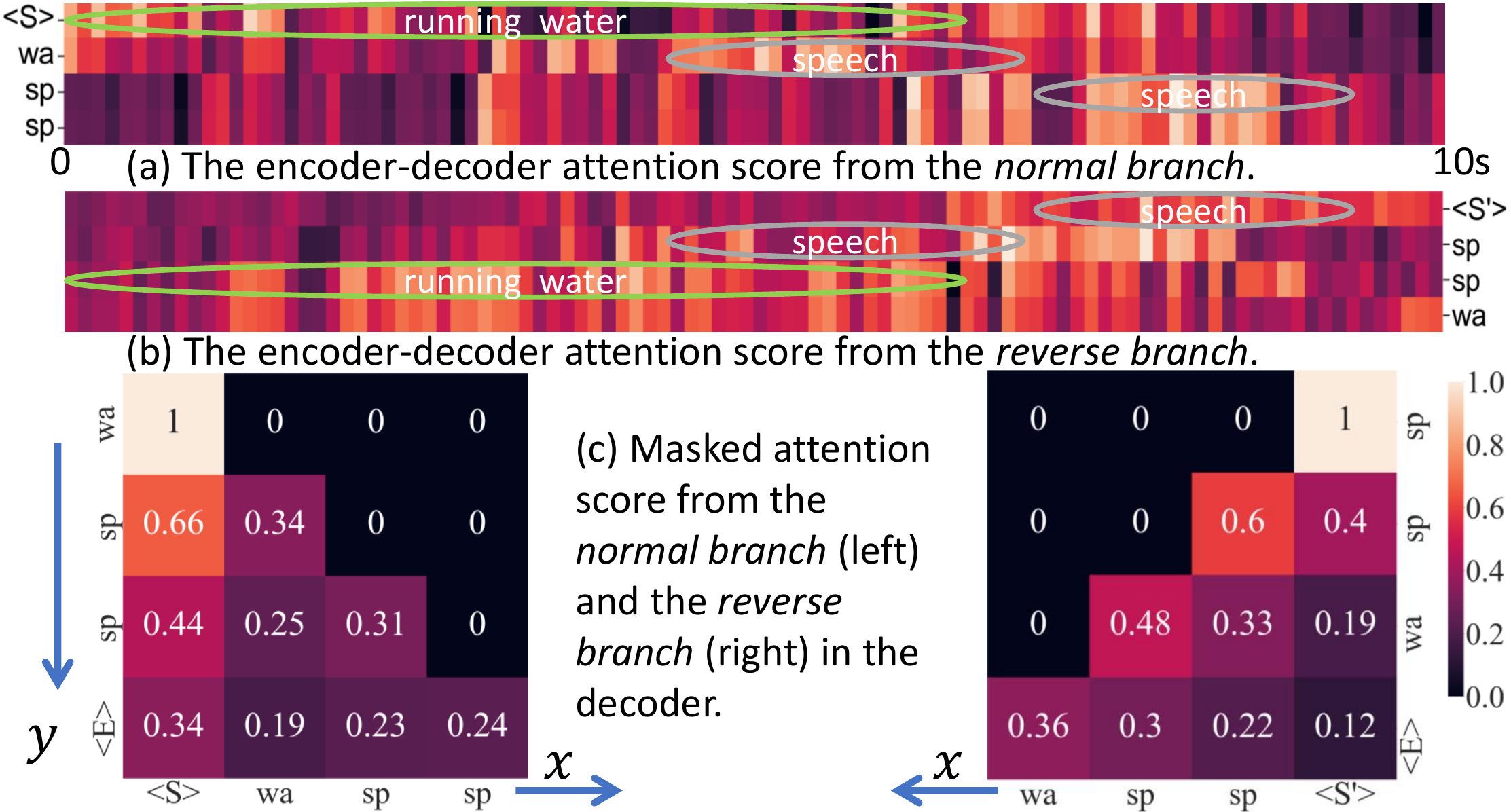}}
	\caption{\small{Attention in GCT. 
In subgraph (c),  the x-axis is each event predicted in an autoregressive way, the y-axis is the reference event.}}
	\label{att}
\end{figure}

\textbf{Case study.}
There is often overlap among various events in polyphonic audio datasets. If one event is almost covered by another, can GCT identify it?
By visualizing the attention distribution,  Fig.~\ref{att} illustrates how GCT recognizes event sequences based on start tokens combined with frame-level acoustic representations from the encoder.
In Fig.~\ref{att} (a) and (b), the x-axis represents the number of audio frames in the encoder, and the y-axis is the event token for each input step in the decoder.
GCT's attention to the audio clip varies after different start tokens are given as input.
For the normal branch of input \S, GCT's attention goes forward from the beginning of the audio clip, focusing on acoustic representations and cues of the next event in turn.
For the reverse order branch of input \Sprime, GCT analyses from the end of the audio clip backward. 
Even though most of the first \textit{speech} (sp) event overlaps with the continuous \textit{running water} (wa) event, GCT can still use the contextual information to effectively identify the covered \textit{speech}.
Fig.~\ref{att} (c) clearly shows how much GCT pays attention to different existing cues in each inference step.
After combining the fine-grained audio representations in Fig.~\ref{att} (a) and (b), under the guidance of \S and \Sprime, the normal and reverse branches of GCT infer event sequences in turn.
The inferred sequences match the corresponding labels consistently, which means that GCT is good at exploiting event context to identify event sequences.

\label{ssec:figure-att}

\vspace{-0.35cm}
\section{CONCLUSION}
\label{sec:CONCLUSION}

\vspace{-0.3cm}
To improve cTransformer in structure and inference, we propose a gated contextual Transformer (GCT) with GCMLP and FBI for SAT.
To gain insight into the role of GCT modules, we study the contribution of different modules to GCT with the help of pretrained weights.
GCT performs well in event classification and sequence modeling, as compared with CTC and Transformer based models. 
Finally, we illustrate the potential of GCT with a visualization of a typical polyphonic audio sample.
Future work will explore GCT on more datasets.

\vfill\pagebreak

\label{sec:refs}

\bibliographystyle{IEEEbib}
\bibliography{Template}

\end{document}